\documentclass[%
 preprint,
 amsmath,amssymb,
]{revtex4-1}

\usepackage{graphicx}
\usepackage{dcolumn}
\usepackage{bm}

\usepackage{graphicx,kantlipsum,setspace}

\begin{document}
%

\title{Inference of hopping rates of anisotropic random walk on a 2D lattice via covariance-based estimators of diffusion parameters}

\author{Masanori Mishima}
 \affiliation{Centre for Mechanochemical Cell Biology, Division of Biomedical Sciences, Warwick Medical School, University of Warwick
 }

\begin{abstract}
Traditionally, time-development of the mean square displacement has been employed to determine the diffusion coefficient from the trajectories of single particles. However, this approach is sensitive to the noise and the motion blur upon image acquisition. Recently, Vestergaard et al.  has proposed a novel method based on the covariance between the shifted displacement series.  This approach gives a more robust estimator of the diffusion coefficient of one-dimensional diffusion without bias, i.e., when mean velocity is zero. Here, we extend this approach to a potentially biased random walk on a two-dimensional lattice. First, we describe the relationship between the  hopping rates to the eight adjacent sites and the time development of the higher-order moments of the stochastic two-dimensional displacements. Then, we derive the covariance-based estimators for these higher-order moments. Numerical simulations confirmed that the procedure presented here allows inference of the stochastic hopping rates from two-dimensional trajectory data with location error and motion blur.

\end{abstract}

\maketitle

\tableofcontents

\section{\label{sec:level1}Introduction}
The forward and backward hopping rates of a random walk on a one-dimensional lattice are linked to the average velocity ($v$) and diffusion coefficient ($D$). The latter macroscopic parameters can be estimated from the trajectory of the random walk, i.e., locations on the lattice measured with a regular time interval. Fitting of the mean square displacements (MSD) vs time lag data with $ MSD=\sigma^2+2D\tau +(v\tau)^2$, where $\sigma$ is the localization noise, has been widely used to estimate the macroscopic parameters. However, it is known that there are practical issues such as the optimization of the range of the time lags and the influence of motion blurs when this approach is applied to the single particle dynamics of a biological molecule \cite{Michalet:2010hd, Berglund:2010ff, Michalet:2012do}. Recently, a novel approach based on the covariance between the displacements between adjacent time points (covariance-based estimator, CVE) has been proposed to be a more simple, accurate and robust alternative to estimate the diffusion coefficient of a unbiased random walk \cite{Vestergaard:2016ko,Vestergaard:2015jx,Vestergaard:2014bk}.

Here, we extend the CVE-based approach to the random walk on a two-dimensional (2D) lattice, in which a particle takes stochastic hops to the eight surrounding sites with distinct rates. We start with the well-known relation between velocity and the diffusion coefficient and the 1D hopping rates and the CVE of the diffusion coefficient in this case. We then show that in a 2D random walk, x- and y-velocities and diffusion coefficients plus four (higher-order) co-moments of the observed two-dimensional displacement series are linked to the eight hopping rates. The procedure for calculating the CVEs for the macroscopic parameters, i.e., the co-moments of the 2D displacements, is provided. This allows us to infer the eight hopping rates from the trajectories of a 2D random walk even with temporal and spatial resolutions at which individual hopping events can't be captured.






\section{\label{sec:level2}Results}
\subsection{\label{sec:level2a}CVE-based approach for 1D diffusion with drift}
\begin{figure*}[htb!]
\includegraphics[width=0.6\textwidth, angle=0]{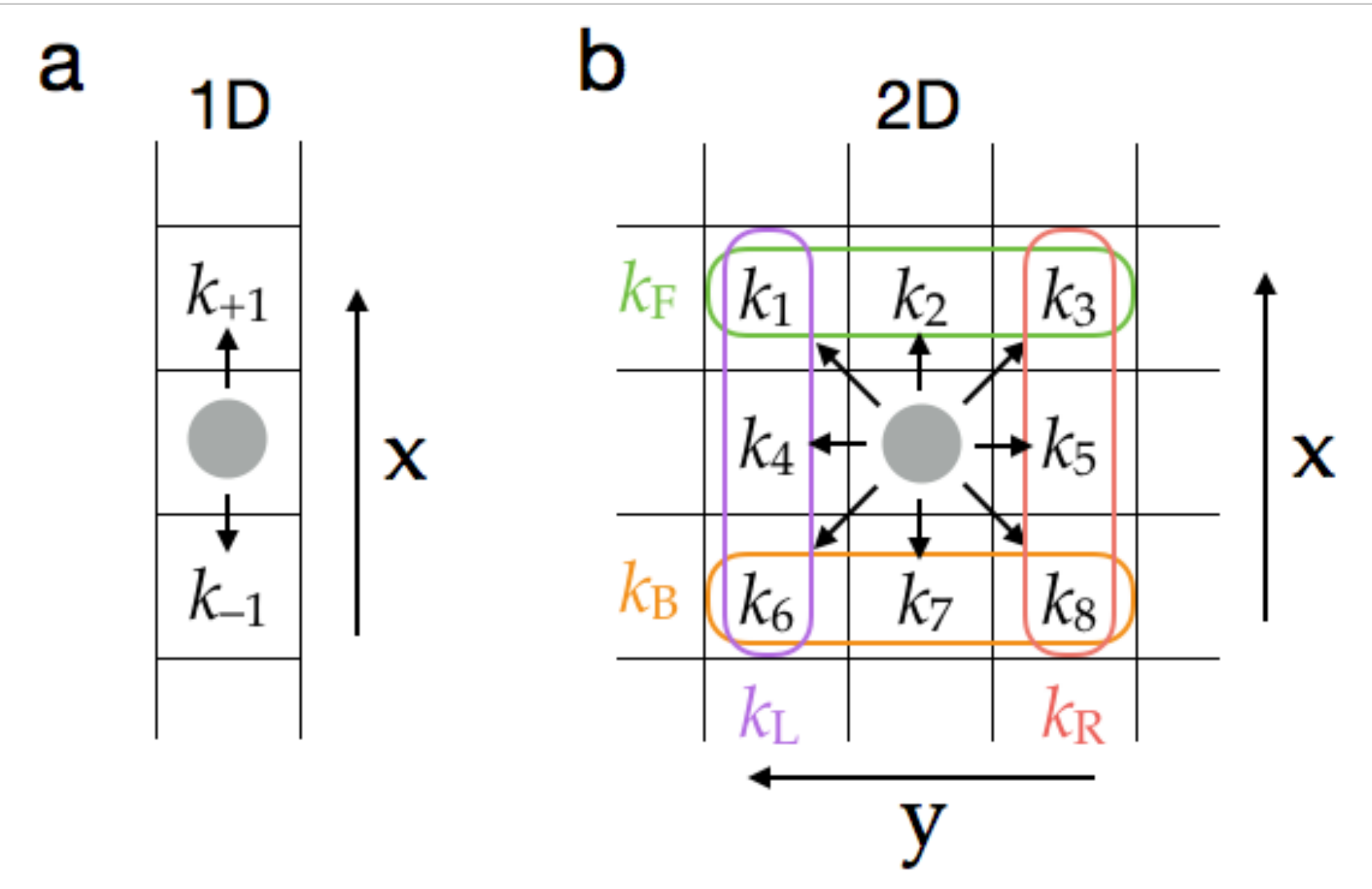}
\caption{\label{fig:stepping} Hopping on 1D (a) and 2D (b) lattices. \newline
The gray dot represents the current position on the lattice. It takes a stochastic hop to the 2 and 8 nearby sites at the indicated rates ($k_{+1}$ and $k_{-1}$ on the 1D lattice (a) and $k_1 \sim k_8 $ on the 2D lattice (b)). $k_F=k_1+k_2+k_3$, $k_B=k_6+k_7+k_8$, $k_L=k_1+k_4+k_6$, and $k_R=k_3+k_5+k_8$ are the rates of hopping towards forward, backward, left and right, respectively.}
\end{figure*}

Here we consider a potentially biased random walk $X(t)$ on a 1D lattice with the grid size, $a$\ (Fig.~\ref{fig:stepping}a). Let $k_{+1}$ and $k_{-1}$ be the stochastic forward and backward hopping rates , respectively (we can assume $k_{+1}>k_{-1}>0$ without loss of generality). $X(t)$  is the sum of $x_i$, the displacements in a small fraction of time $\delta t = t/n$, i.e.,
$ X(t) = x_1 + x_2 + \ldots +x_n $.

Since $x_i$ are indistinguishable but independent of each other, the ensemble averages of $x_i$
\begin{eqnarray*}
    \langle x_1 \rangle = \langle x_2 \rangle = \ldots = \langle x_n \rangle \equiv \langle x \rangle \\
    \langle x_1^2 \rangle = \langle x_2^2 \rangle = \ldots = \langle x_n^2 \rangle  \equiv \langle x^2 \rangle 
\end{eqnarray*}
where  $\langle x \rangle$ and $\langle x^2 \rangle$ are the mean displacement and the mean square displacement of $X(t)$ in $\delta t$, respectively. In general, $\langle x_i^2 \rangle \neq \langle x_i \rangle^2 $, while for $i \neq j$, $\langle x_ix_j \rangle = \langle x_i \rangle \langle x_j \rangle $. The ensemble average of $X(t)$ is 
\begin{eqnarray}
    \langle X(t) \rangle &=& \langle x_1 + x_2 + \ldots +x_n \rangle = n\langle x \rangle \nonumber \\
        &=& n\bigl[ (+a)\times k_{+1}\delta t + (-a)\times k_{-1}\delta t\bigr] = a(k_{+1} - k_{-1}) \cdot n\delta t  \nonumber \\
        &=& a(k_{+1} - k_{-1})t \nonumber
\end{eqnarray}

With $\langle X(t)^2 \rangle =  \langle (x_1 + x_2 + \ldots +x_n)^2 \rangle = \langle \sum_i x_i^2 + \sum_j \sum_{i, i \neq j} x_i x_j \rangle = n \langle x^2 \rangle + n(n-1) \langle x \rangle^2$, the variance of $X(t)$ is
\begin{eqnarray}
    \mathrm{var} (X(t)) &=&\langle (X(t)-\langle X(t) \rangle)^2 \rangle =\langle X(t)^2 \rangle - \langle X(t) \rangle ^2 \nonumber \\
    &=&  n \langle x^2 \rangle + n(n-1) \langle x \rangle^2 - (n\langle x \rangle)^2 = n \langle x^2 \rangle - n \langle x \rangle^2 \nonumber \\
    &=& n\bigl[ (+a)^2\times k_{+1}\delta t + (-a)^2\times k_{-1}\delta t\bigr] - n \bigl[ (+a)\times k_{+1}\delta t + (-a)\times k_{-1}\delta t\bigr]^2 \nonumber \\
   &=& a^2(k_{+1} + k_{-1}) \cdot n\delta t  -   a^2(k_{+1} - k_{-1})^2 (n\delta t)\delta t \nonumber \\
    &\rightarrow & a^2(k_{+1} + k_{-1})t \ \ (\delta t \rightarrow 0)\nonumber
\end{eqnarray}
Thus, the velocity of the constant drift and the diffusion coefficient, $v$ and $D$, respectively, are linked to the hopping rates
\begin{eqnarray}
    v  = a (k_{+1} - k_{-1}) \\
    2D = a^2 (k_{+1} + k_{-1}) 
\end{eqnarray}
As is well known, this indicates that we can infer the stochastic hopping rates by determining the macroscopic parameters, $v$ and $D$. Hereafter, for simplicity, we omit $a$ by assuming that $X(t)$ is a dimensionless value measured with $a$ as the unit, i.e., the physical location on the lattice is $X(t)\cdot a$. Thus,
\begin{eqnarray}
    v  = k_{+1} - k_{-1} \\
    2D = k_{+1} + k_{-1}
\end{eqnarray}

In typical single molecule/particle observations, we can't determine the true coordinate $X(t)$ as a continuous function. We can only measure the positions at a discrete time points $t=0, \Delta t, 2\Delta t, \dots , n \Delta t$ and the measured values $X_0, X_1, X_2, \dots X_n $ suffer from the motion blur due to the movement during image capturing and the error in determining the position of the molecule/particle by image analysis. Under this circumstance, $X_k$ is related to the true $X(t)$ as
\begin{equation}
    X_k = \int_0^{\Delta t} s(t) X(t + (k-1)\Delta t) dt + \varepsilon_k
\end{equation}
where $s(t)$ defines the state of the shutter ($s(t)=0$ means closed shutter and $s(t)>0$ means open shutter, for normalization $\int_0^{\Delta t} s(t) dt =1$) and $\varepsilon_k$ the Gaussian error in localization by image analysis ($\langle \varepsilon_i\rangle =0$ and $\langle \varepsilon_i^2\rangle =\sigma^2$ represents the precision of the measurement when the target is immobile) \cite{Berglund:2010ff}. The traditional approaches to determine $v$ and $D$, for example, by fitting a quadratic curve to the mean square displacement data, suffer from complications arising from non-zero $s(t)$ and $\varepsilon$. However, for unbiased diffusion (i.e. $v=0$), it has recently been shown \cite{Vestergaard:2014bk, Vestergaard:2015jx, Vestergaard:2016ko} that a combination of the adjacent displacements $\Delta X_k = X_{k+1} - X_k$ and $\Delta X_{k+1} = X_{k+2} - X_{k+1}$ cancels out the terms containing $s(t)$ and $\varepsilon$ and results in a simple relation
\begin{equation}
    \langle \Delta X_k^2 \rangle + 2 \langle \Delta X_{k+1}\cdot \Delta X_k \rangle = 2D \Delta t.
\end{equation}
$D$ calculated with this relation provides a more reliable estimator of the diffusion coefficients.

The above relation doesn't hold in the presence of a bias in the hopping rates. However, if we consider $Z(t) = X(t) - vt$, the deviation from the constant drift, and its observed counterpart $Z_k = X_k- v\cdot k\Delta t$, we obtain $\langle Z_k \rangle = 0$, and thus the series of its displacements $\Delta Z_k = Z_{k+1} - Z_k = \Delta X_k - v\Delta t$ satisfy 
\begin{equation}
    \langle \Delta Z_k^2 \rangle + 2 \langle \Delta Z_{k+1}\cdot \Delta Z_k \rangle = 2D \Delta t.
\end{equation}
Since $\langle (X(t) - \langle X(t) \rangle)^2 \rangle =  \langle (Z(t) + vt - \langle Z(t) +vt\rangle)^2 \rangle =  \langle (Z(t) - \langle Z(t) \rangle)^2 \rangle$, the value of $D$ determined with this formula gives the diffusion coefficient of the biased random walk $X(t)$. 

\subsection{\label{sec:level2b} Biased random walk on a 2D lattice: hopping rates and co-moments of the displacements}
Here we discuss a biased random walk $(X(t), Y(t))$ on a 2D lattice measured with the x- and y-grid size $a$ and $b$, respectively. Hopping can occur to the eight surrounding sites with the distinct rates $k_1, k_2, k_3, k_4, k_5, k_6, k_7, k_8$. Let the velocities and diffusion coefficients of the movement along the grid axes be $v_x$ and  $v_y$, and $D_x$, and $D_y$, respectively. Then 
\begin{eqnarray*}
    \langle X(t) \rangle = v_x t = (k_F-k_B) t = (k_1 + k_2 + k_3 - k_6 - k_7 - k_8) t \\
    \langle ( X(t) - \langle X(t) \rangle)^2 \rangle = 2 D_x t =  (k_F+k_B) t = (k_1 + k_2 + k_3 + k_6 + k_7 + k_8) t\\
    \langle Y(t) \rangle = v_y t = (k_L-k_R) t= (k_1 + k_4 + k_6 - k_3 - k_5 - k_8) t\\
    \langle ( Y(t) - \langle Y(t) \rangle)^2 \rangle = 2 D_y t = (k_L+k_R) t=   (k_1 + k_4 + k_6 + k_3 + k_5 + k_8) t
\end{eqnarray*}
as we discussed in the above section.

Here, we consider the covariance between $X(t)$ and $Y(t)$.
\begin{eqnarray*}
   && \mathrm{cov}(X, Y) = \langle (X - \langle X \rangle)(Y - \langle Y \rangle) \rangle = \langle XY \rangle - \langle X\rangle \langle Y \rangle \\
   &=& \langle (x_1+x_2+ \ldots + x_n)(y_1+y_2+ \ldots +y_n) \rangle - \langle x_1 + x_2 + \ldots + x_n \rangle\langle y_1 + y_2 + \ldots + y_n \rangle \\
   &=& \langle \sum_{i=1}^n x_i y_i + \sum_{i\neq j}x_i y_j  \rangle - n\langle x \rangle \cdot n \langle y \rangle \\
  &=& n \langle xy \rangle + n(n-1) \langle x\rangle \langle y \rangle - n^2 \langle x \rangle \langle y \rangle \\
   &=& n \langle xy \rangle -  n \langle x\rangle \langle y \rangle \\
   &=& n \cdot \bigl[ (+1)(+1)\cdot k_1\delta t + (-1)(+1)\cdot k_3\delta t + (+1)(-1)\cdot k_6\delta t + (-1)(-1)\cdot k_8\delta t \bigr] \\&&\ \ \ \  - n \cdot (k_F - k_B)\delta t \cdot (k_L - k_R)\delta t \\
   &=& (k_1-k_3 -k_6+k_8) n\delta t - (k_F - k_B) (k_L - k_R) n\delta t \cdot \delta t \\
   &\rightarrow& (k_1-k_3 -k_6+k_8)t\ \ (\delta t \rightarrow 0)
\end{eqnarray*}

We move on to higher-order moments. 
Since
\begin{eqnarray}
    \langle X^2 Y \rangle &=& \Bigl\langle \sum_k \sum_j \sum_i x_i x_j y_k \Bigr\rangle \nonumber \\
   &=& n  \langle x^2 y \rangle +n(n-1) \langle x^2\rangle \langle y \rangle + 2n(n-1)\langle x \rangle \langle xy \rangle + n(n-1)(n-2) \langle x \rangle ^2 \langle y \rangle \nonumber , 
\end{eqnarray}
we get
\begin{eqnarray*}
   && \bigl\langle (X - \langle X \rangle)^2( Y - \langle Y \rangle) \bigr\rangle = \langle X^2 Y \rangle - \langle X^2 \rangle \langle Y \rangle -2 \langle X \rangle (\langle XY \rangle - \langle X \rangle \langle Y \rangle )  \\
   &=& n \langle x^2 y \rangle - n \langle x^2 \rangle \langle y \rangle - 2n \langle x\rangle \langle xy \rangle +2n \langle x \rangle^2 \langle y \rangle \\
  & =& (k_1 - k_3 +k_6 - k_8) n\delta t - \Bigl[(k_F + k_B)(k_L - k_R) + 2(k_F - k_B)(k_1 - k_3 -k_6 + k_8)\Bigr] n \delta t \cdot \delta t  \\
 && \ \ \ \ \ \ \ \ \ \ +2(k_F - k_B)^2 (k_L - k_R) n\delta t \cdot \delta t^2 \\
  & \rightarrow & (k_1 - k_3 +k_6 - k_8)t\ \ (\delta t \rightarrow 0).
\end{eqnarray*}

Similarly, 
\begin{equation*}
     \bigl\langle (X - \langle X \rangle)( Y - \langle Y \rangle)^2 \bigr\rangle \rightarrow (k_1+k_3 - k_6 -k_8) t \ \ (\delta t \rightarrow 0)
\end{equation*}

Finally, using
\begin{eqnarray*}
  && \langle X^2 Y^2 \rangle = \Bigl \langle \sum_l \sum_k \sum_j \sum_i x_i x_j y_k y_l \Bigr \rangle \\
   &=& n\langle x^2 y^2 \rangle + 2n(n-1) \langle x \rangle \langle x y^2 \rangle + 2n(n-1) \langle y \rangle \langle x^2 y \rangle + 2n(n-1)\langle xy\rangle^2 \nonumber \\ &&\ \ \ \ + n(n-1)\langle x^2 \rangle \langle y^2 \rangle + n(n-1)(n-2) \langle x^2 \rangle \langle y \rangle^2 +4n(n-1)(n-2) \langle xy \rangle \langle x\rangle \langle y \rangle \nonumber \\
   &&\ \ \ \ \ \ \ \ \  + n(n-1)(n-2)\langle x \rangle^2 \langle y^2 \rangle + n(n-1)(n-2)(n-3) \langle x \rangle^2 \langle y \rangle^2 \nonumber 
\end{eqnarray*}
we get 
\begin{eqnarray*}
 &&\Bigl\langle (X- \langle X \rangle)^2  (Y- \langle Y \rangle)^2 \Bigr\rangle\\
 &=& \langle X^2 Y^2 \rangle - 2\langle X^2 Y \rangle \langle Y \rangle -2\langle X Y^2 \rangle \langle X \rangle  + \langle X^2 \rangle \langle Y \rangle^2  + 4\langle XY \rangle\langle X \rangle\langle Y \rangle +\langle Y^2 \rangle \langle X \rangle^2 -3 \langle X\rangle^2  \langle Y\rangle^2  \\
 &=& n \langle x^2 y^2 \rangle + 2n(n-1) \langle xy \rangle ^2 + n(n-1)\langle x^2\rangle \langle y^2 \rangle  \\
 &&\ \ \ \ \ \ \ - 2n \bigl[ \langle x^2 y \rangle \langle y \rangle+\langle xy^2 \rangle \langle x \rangle \bigr] - n(n-2) \bigl[\langle x^2\rangle \langle y \rangle^2 +4 \langle xy \rangle \langle x \rangle \langle y \rangle + \langle y^2\rangle \langle x \rangle^2 -3\langle x \rangle^2 \langle y \rangle^2\bigr] \\
   &\rightarrow & (k_1+k_3+k_6+k_8) t + 2(At)^2 + 2D_x t \cdot 2D_y t \ \ (\delta t \rightarrow 0)
\end{eqnarray*}

In summary, we have obtained a set of formulae that relate macroscopic observations to the microscopic hopping rates:
\begin{eqnarray}
    \label{def_pars1}
        \langle X \rangle &=& v_x t \\
    \label{def_pars2}
        \langle Y \rangle &=& v_y t \\
    \label{def_pars3}
        \langle (X - \langle X \rangle)^2 \rangle &=& 2D_x t \\
    \label{def_pars4}
        \langle (Y - \langle Y \rangle)^2 \rangle &=& 2D_y t \\
    \label{def_pars5}
        \langle (X - \langle X \rangle)(Y - \langle Y \rangle) \rangle &=& A t \\
    \label{def_pars6}
        \langle (X - \langle X \rangle)^2(Y - \langle Y \rangle) \rangle &=& B t \\
    \label{def_pars7}
        \langle (X - \langle X \rangle)(Y - \langle Y \rangle)^2 \rangle &=& C t \\
    \label{def_pars8}
        \langle (X - \langle X \rangle)^2(Y - \langle Y \rangle)^2 \rangle \nonumber   - 2 \langle (X - \langle X \rangle)(Y - \langle Y \rangle)  \rangle^2\ \ \ \ \ \ \ \ &&\\
 - \langle (X - \langle X \rangle)^2 \rangle \langle (Y - \langle Y \rangle)^2 \rangle &=& E t
\end{eqnarray}
where
\begin{equation}
\label{pars_rates_relation}
\begin{pmatrix}
v_x \\
v_y \\
2D_x \\
2D_y \\
A \\
B \\
C \\
E \\
\end{pmatrix} 
=
\begin{pmatrix}
1  &  1  &  1  &  0  &  0  & -1  & -1  & -1 \\
1  &  0  & -1  &  1  & -1  &  1  &  0  & -1 \\
1  &  1  &  1  &  0  &  0  &  1  &  1  &  1 \\
1  &  0  &  1  &  1  &  1  &  1  &  0  &  1 \\
1  &  0  & -1  &  0  &  0  & -1  &  0  &  1 \\
1  &  0  & -1  &  0  &  0  &  1  &  0  & -1 \\
1  &  0  &  1  &  0  &  0  & -1  &  0  & -1 \\
1  &  0  &  1  &  0  &  0  &  1  &  0  &  1 \\
\end{pmatrix} 
\begin{pmatrix}
k_1 \\
k_2 \\
k_3 \\
k_4 \\
k_5 \\
k_6 \\
k_7 \\
k_8 \\
\end{pmatrix} 
\end{equation}

\begin{figure*}[htb!]
\includegraphics[width=0.8\textwidth, angle=0]{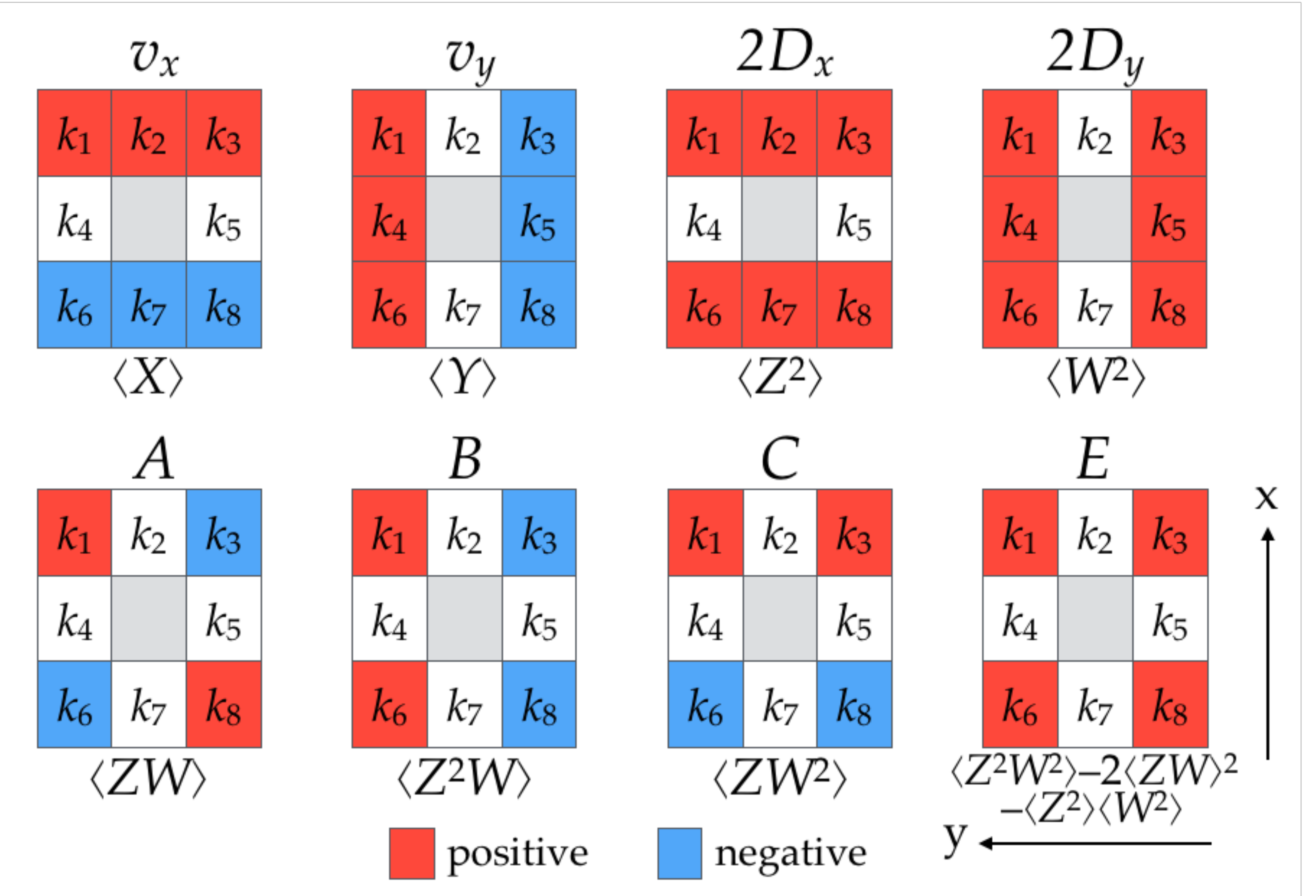}
\caption{\label{fig:rates_pars} Visual presentation of the relationships between the macroscopic parameters ($v_x, v_y, D_x, D_y, A, B, C\  \mathrm{and\ } E$), which are the proportional coefficients of the time development of $\langle X \rangle \mathrm{\ and\ } \langle Y \rangle$ ($v_x$ and $v_y$) or the moments of the drift-adjusted displacements, $Z = X -v_x t, W=Y-v_y t$ ($D_x, D_y, A, B, C\ \mathrm{and\ } E$)(\ref{def_pars1} to \ref{def_pars8}), and the microscopic hopping rates ($k_1, k_2, \ldots, k_8$) on a 2D lattice. The red and blue colors on the grid indicate the signs (positive and negative, respectively) of the rates of hopping from the current position (gray) to the corresponding sites in assembling the macroscopic parameters (\ref{pars_rates_relation}).}
\end{figure*}

\subsection{\label{sec:level2c}Covariance-based estimator for $A$}
Based on formula \ref{def_pars1} and \ref{def_pars2}, the velocities of constant drift, $v_x$ and $v_y$, can be estimated with
\begin{eqnarray}
\label{cve_vx}
    \langle \Delta X \rangle = v_x \Delta t \\
\label{cve_vy}
    \langle \Delta Y \rangle = v_y \Delta t
\end{eqnarray}
As discussed in section A, drift-adjusted displacements $\Delta Z_k = \Delta X_k - v_x \Delta t$ and $\Delta W_k = \Delta Y_k - v_y \Delta t$ based on $Z(t)=X(t) - v_x t$ and $W(t)=Y(t) - v_y t$ give the diffusion coefficients along x- and y-axes, $D_x$ and $D_y$, respectively.
\begin{eqnarray}
\label{cve_2Dx}
    \langle \Delta Z_k^2 \rangle + 2\langle \Delta Z_{k+1} \Delta Z_k \rangle = 2D_x \Delta t \\
\label{cve_2Dy}
    \langle \Delta W_k^2 \rangle + 2\langle \Delta W_{k+1} \Delta W_k \rangle = 2D_y \Delta t 
\end{eqnarray}
Here we derive analogous covariance-based estimators of $A$, $B$, $C$, and $E$. Considering the symmetry between $Z(t)$ and $W(t)$, we guess the formula for $A$ as
\begin{equation*}
   \sum_{\substack{\alpha, \beta=0\ \mathrm{or}\ 1 \\ \mathrm{not\ } \alpha=\beta=1}} \langle \Delta Z_{k+\alpha} \Delta W_{k+\beta}\rangle = \langle \Delta Z_k \Delta W_k \rangle + \langle \Delta Z_{k+1} \Delta W_k\rangle + \langle \Delta Z_k \Delta W_{k+1}\rangle = A \Delta t   
\end{equation*}
We shall now prove this.

With
\begin{eqnarray}
    \Delta Z_{k+\alpha} &=& \int_0^{\Delta t} s(t) \Bigl[
        Z(t+(k+\alpha)\Delta t) - Z(t+(k+\alpha-1)\Delta t)
    \Bigr] dt +(\varepsilon_{k+\alpha+1} - \varepsilon_{k+\alpha}) \nonumber \\
    \Delta W_{k+\beta} &=& \int_0^{\Delta t} s(t') \Bigl[
        W(t'+(k+\beta)\Delta t) - W(t'+(k+\beta-1)\Delta t) \nonumber
    \Bigr] dt' +(\varepsilon'_{k+\beta+1} - \varepsilon'_{k+\beta})  
\end{eqnarray}

\begin{eqnarray*}
    &&\langle \Delta Z_{k+\alpha} \Delta W_{k+\beta} \rangle  \\
    &=& \int_0^{\Delta t} \int_0^{\Delta t} s(t)s(t') \Bigl \langle
    \Bigl( Z(t+(k+\alpha)\Delta t) -   Z(t+(k+\alpha-1)\Delta t)\Bigr)  \\
   &&\ \ \times \Bigl( W(t'+(k+\beta)\Delta t) -   W(t'+(k+\beta-1)\Delta t) \Bigr)
   \Bigr \rangle dt dt' +\langle (\varepsilon_{k+\alpha+1} - \varepsilon_{k+\alpha})(\varepsilon'_{k+\beta+1} - \varepsilon'_{k+\beta})  \rangle 
\end{eqnarray*}
Since $(Z(t), W(t))$ is a random walk that fulfils $\langle Z(t)W(t) \rangle =At$,  $\langle Z(t)W(t') \rangle =A\cdot \mathrm{min}(t, t')$. Using this, the factor to be integrated in the first term of $\langle \Delta Z_{k+\alpha} \Delta W_{k+\beta} \rangle $ is evaluated to be
\begin{eqnarray*}
&&\Bigl \langle
    \Bigl( Z(t+(k+\alpha)\Delta t) -   Z(t+(k+\alpha-1)\Delta t)\Bigr)   \times \Bigl( W(t'+(k+\beta)\Delta t) -   W(t'+(k+\beta-1)\Delta t) \Bigr) \Bigr \rangle \\
    &=& \bigl \langle Z(t+(k+\alpha)\Delta t) W(t'+(k+\beta)\Delta t) \bigr \rangle  - \bigl \langle Z(t+(k+\alpha)\Delta t)  W(t'+(k+\beta-1)\Delta t) \bigr \rangle\\
   && - \bigl \langle  Z(t+(k+\alpha-1)\Delta t) W(t'+(k+\beta)\Delta t) \bigr \rangle + \bigl \langle  Z(t+(k+\alpha-1)\Delta t) W(t'+(k+\beta-1)\Delta t) \bigr \rangle \\
   &=& A \Bigl[\{ k\Delta t + \mathrm{min}(t + \alpha \Delta t, t'+\beta \Delta t)\} - \{k\Delta t + \mathrm{min}(t + \alpha \Delta t, t'+ (\beta-1) \Delta t)\}  \\
&&\ \ \ \ \ \ \ \ \ \ \ \ \ \ \ \ \ \ \ \ \ -  \{k\Delta t +\mathrm{min}(t + (\alpha-1) \Delta t, t'+ \beta \Delta t)\}  + \{(k-1)\Delta t + \mathrm{min}(t + \alpha \Delta t, t'+\beta \Delta t) \} \Bigr]\\
   &=& A \Bigl[ \mathrm{min}(t + \alpha \Delta t, t'+\beta \Delta t) - \mathrm{min}(t + \alpha \Delta t, t'+ (\beta-1) \Delta t)  \\
&&\ \ \ \ \ \ \ \ \ \ \ \ \ \ \ \ \ \ \ \ \ \ \ \ \ \ \ \ \  - \mathrm{min}(t + (\alpha-1) \Delta t, t'+ \beta \Delta t)  + \mathrm{min}(t + \alpha \Delta t, t'+\beta \Delta t)  -\Delta t \Bigr] 
\end{eqnarray*}
Here, considering $ 0 \leq t  \leq \Delta t, 0 \leq t' \leq \Delta t $, the factor within the above brackets 
\begin{eqnarray*}
g(\alpha, \beta, t, t') &=&
\mathrm{min}(t + \alpha \Delta t, t'+\beta \Delta t) - \mathrm{min}(t + \alpha \Delta t, t'+ (\beta-1) \Delta t)  \\
&&\ \ \ \ \ \ \ \ \ \ - \mathrm{min}(t + (\alpha-1) \Delta t, t'+ \beta \Delta t)  + \mathrm{min}(t + \alpha \Delta t, t'+\beta \Delta t)  -\Delta t
\end{eqnarray*}
is calculated to be 
\begin{eqnarray*}
    g(0, 0, t, t^{\prime}) &=&  \Delta t - \bigl(t + t^{\prime} - 2\cdot \mathrm{min}(t, t^{\prime}) \bigr)\\
    g(1, 0, t, t^{\prime}) &=& t^{\prime} - \mathrm{min}(t, t^{\prime}) \\
    g(0, 1, t, t^{\prime}) &=& t - \mathrm{min}(t, t^{\prime})
\end{eqnarray*}
Thus, $g(0, 0, t, t') + g(1, 0, t, t') + g(0, 1, t, t') = \Delta t$.

The x- and y-localization errors at the same time point are not independent. With Kronecker's delta $\delta (\alpha, \beta) = 1\ (\alpha=\beta)\ \mathrm{or}\ 0\ (\alpha \neq \beta)$ ,
\begin{eqnarray*}
&&\Bigl\langle ( \varepsilon_{k+\alpha+1} - \varepsilon_{k+\alpha})  ( \varepsilon'_{k+\beta+1} - \varepsilon'_{k+\beta}) \Bigr\rangle \\
&=& \langle \varepsilon_{k+\alpha+1}\varepsilon'_{k+\beta+1}  \rangle - \langle   \varepsilon_{k+\alpha+1} \varepsilon'_{k+\beta} \rangle - \langle \varepsilon_{k+\alpha} \varepsilon'_{k+\beta+1}\rangle + \langle \varepsilon_{k+\alpha} \varepsilon'_{k+\beta} \rangle \\
&=& \sigma_{xy}^2 (\delta(\alpha, \beta) - \delta(\alpha+1, \beta) - \delta(\alpha, \beta+1) + \delta(\alpha, \beta))\\
&=& \sigma_{xy}^2 (2 \delta(\alpha, \beta) - \delta(\alpha+1, \beta) - \delta(\alpha, \beta+1)) = \sigma_{xy}^2 \Omega(\alpha, \beta),
\end{eqnarray*}
where we define $\Omega(\alpha, \beta)=2 \delta(\alpha, \beta) - \delta(\alpha+1, \beta) - \delta(\alpha, \beta+1)$.
Thus, $ \Omega(0, 0) + \Omega(0, 1)+ \Omega(1, 0) = (2\cdot 1 - 0 - 0) + ( 2\cdot 0 - 1 - 0) + (2\cdot 0 - 0 - 1) =0$.
    
Finally, we get
\begin{eqnarray}
\label{cve_A}
&&\sum_{\substack{\alpha, \beta=0\ \mathrm{or}\ 1 \\ \mathrm{not\ } \alpha=\beta=1}} \langle \Delta Z_{k+\alpha} \Delta W_{k+\beta}\rangle \nonumber \\
&=& \int_0^{\Delta t} \int_0^{\Delta t} s(t)s(t') A \cdot \bigl[g(0, 0, t, t') + g(1, 0, t, t') + g(0, 1, t, t')\bigr] dt dt' \nonumber \\
&&\ \ \ \ \ \ \ \ \ \ \ \ \ \ \ \ \ \ \ \ \ \ \ \ \ \ \ \ \ \ \ \ \ \ \ \ \ \ \ \ \ \ \ \ \ \ \ \ \ \ \ \ \ \ \ \ +\sigma_{xy}^2 \bigl[\Omega(0, 0) + \Omega(1, 0)+ \Omega(0, 1)\bigr] \nonumber \\
&=& A \Delta t
\end{eqnarray}

\subsection{\label{sec:level2d}Covariance-based estimators for $B$ and  $C$ }
The above results imply that for random walks $U(s), V(t),\ldots$, if a function $f(x, y,\ldots )$ exists such that $ f(U(s), V(t), \dots) = K \cdot \mathrm{min}(s, t, \dots)$ for some constant $K$ (for example, for $f(x, y)=\langle xy\rangle$,  $f(X(s), X(t))=2D_x  \cdot \mathrm{min}(s,t)$,  and $f(X(s), Y(t))=A  \cdot \mathrm{min}(s, t)$), then the equation below holds
\begin{equation}
\label{cve_recipe}
    \sum_{\substack{\alpha, \beta, \ldots =0\ \mathrm{or\ }1 \\ \mathrm{not\ }\alpha=\beta=\ldots = 1}} f(\Delta U_{k+\alpha}, \Delta V_{k+\beta},\dots) = K\Delta t
\end{equation}
where $ \Delta U_k = U_{k+1} - U_{k}$, $ \Delta V_k = V_{k+1} - V_{k}\ldots$ are the observed displacements (adjusted for the constant drift), providing the general recipe for the CVE for $K$. Below is a proof for the cases of $B$ and $C$.

Since $\langle Z(t)^2 W(t) \rangle = B t$, we choose $ f(x, y, z) = \langle xyz \rangle $.
\begin{eqnarray*}
    && f(\Delta Z_{k+\alpha}, \Delta Z_{k+\beta}, \Delta W_{k+\gamma} ) = \langle \Delta Z_{k+\alpha} \cdot  \Delta Z_{k+\beta}  \cdot \Delta W_{k+\gamma} \rangle \\
  &=&  \Bigl \langle \Bigl(\int_0^{\Delta t} s(t)\Bigl[Z(t + (k + \alpha)\Delta t) - Z(t + (k + \alpha-1)\Delta t)\Bigr] dt +(\varepsilon_{k+\alpha+1}- \varepsilon_{k+\alpha}) \Bigr) \\
  &&\ \ \ \ \ \ \ \times \Bigl( \int_0^{\Delta t} s(t')\Bigl[Z(t' + (k + \beta)\Delta t) - Z(t' + (k + \beta-1)\Delta t)\Bigr] dt' +(\varepsilon_{k+\beta+1}- \varepsilon_{k+\beta}) \Bigr) \\
  &&\ \ \ \ \ \ \ \ \ \ \ \ \ \ \ \ \ \times \Bigl( \int_0^{\Delta t} s(t'')\Bigl[W(t'' + (k + \gamma)\Delta t) - W(t'' + (k + \gamma-1)\Delta t)\Bigr] dt'' +(\delta_{k+\gamma+1}- \delta_{k+\gamma}) \Bigr) \Bigr \rangle \\
  &=& \int_0^{\Delta t} \int_0^{\Delta t} \int_0^{\Delta t} s(t)s(t')s(t'') \\ 
  &&\ \ \ \ \ \ \Bigl \langle
  \Bigl[Z(t + (k + \alpha)\Delta t) - Z(t + (k + \alpha-1)\Delta t)\Bigr] \times  \Bigl[Z(t' + (k + \beta)\Delta t) - Z(t' + (k + \beta-1)\Delta t)\Bigr] \\
  &&\ \ \ \ \ \ \ \ \ \ \ \times   \Bigl[W(t'' + (k + \gamma)\Delta t) - W(t'' + (k + \gamma-1)\Delta t)\Bigr] 
  \Bigr \rangle dt dt' dt'' \\
  &&\ \ \ \ \ \ \ \ \ \ \ \ \ \ \ \ \ \ \ \  + \Bigl \langle (\varepsilon_{k+\alpha+1}- \varepsilon_{k+\alpha}) (\varepsilon_{k+\beta+1}- \varepsilon_{k+\beta}) (\delta_{k+\gamma+1}- \delta_{k+\gamma}) \Bigr \rangle, \\
\end{eqnarray*} 
in which we used $ \langle Z(t+(k+\alpha)\Delta t) - Z(t+(k+\alpha-1)\Delta t) \rangle = 0$, etc. and $\langle \varepsilon_{k+\alpha+1}- \varepsilon_{k+\alpha} \rangle = 0$, etc.

 For unbiased random walks $A(s), B(t), C(u)$ ($\langle A(s) \rangle=\langle B(t) \rangle=\langle C(u) \rangle = 0$, and their displacements are independent for non-overlapping time sections), if we assume $s<t <u$,
\begin{eqnarray}
   &&  \langle A(s) B(t) C(u) \rangle \nonumber = \langle A(s) \rangle \langle (B(t) - B(s))(C(t)-C(s)) \rangle 
    + \langle A(s)C(s)\rangle ( \langle B(t) \rangle - \langle B(s) \rangle) \nonumber \\
   &&\ \ \  + \langle A(s)B(s)\rangle ( \langle C(t) \rangle - \langle C(s) \rangle)
    + \langle A(s) B(s) C(s) \rangle 
\label{wiener_process}
    = \langle A(s) B(s) C(s) \rangle
\end{eqnarray}
In general, with $\tau = \mathrm{min}(s, t, u)$, $    \langle A(s) B(t) C(u) \rangle = \langle A(\tau) B(\tau) C(\tau) \rangle$.
Applying this to $Z(t)$ and $W(t)$ that satisfy $\langle Z(t)^2 W(t) \rangle = B t$, we get $ \langle Z(t)Z(t')W(t'') \rangle = B\cdot \mathrm{min} (t, t', t'') $.
Using this, we evaluate
\begin{eqnarray*}
    &&\Bigl \langle  \Bigl[Z(t + (k + \alpha)\Delta t) - Z(t + (k + \alpha-1)\Delta t)\Bigr] \times  \Bigl[Z(t' + (k + \beta)\Delta t) - Z(t' + (k + \beta-1)\Delta t)\Bigr] \\
    &&\ \ \ \ \ \ \ \ \ \ \ \times  \Bigl[W(t'' + (k + \gamma)\Delta t) - W(t'' + (k + \gamma-1)\Delta t)\Bigr]  \Bigr \rangle \\
  & =& \bigl \langle  Z(t + (k + \alpha)\Delta t) \cdot Z(t' + (k + \beta)\Delta t) \cdot W(t'' + (k + \gamma)\Delta t)   \bigr \rangle \\
  &&\ \  -\bigl \langle  Z(t + (k + \alpha)\Delta t) \cdot Z(t' + (k + \beta)\Delta t) \cdot W(t'' + (k + \gamma-1)\Delta t)   \bigr \rangle \\
   &&\ \  -\bigl \langle  Z(t + (k + \alpha)\Delta t) \cdot Z(t' + (k + \beta-1)\Delta t) \cdot W(t'' + (k + \gamma)\Delta t)   \bigr \rangle \\
    &&\ \  +\bigl \langle  Z(t + (k + \alpha)\Delta t) \cdot Z(t' + (k + \beta-1)\Delta t) \cdot W(t'' + (k + \gamma-1)\Delta t)   \bigr \rangle \\
  &&\ \   -  \bigl \langle  Z(t + (k + \alpha-1)\Delta t) \cdot Z(t' + (k + \beta)\Delta t) \cdot W(t'' + (k + \gamma)\Delta t)   \bigr \rangle \\
  &&\ \  +\bigl \langle  Z(t + (k + \alpha-1)\Delta t) \cdot Z(t' + (k + \beta)\Delta t) \cdot W(t'' + (k + \gamma-1)\Delta t)   \bigr \rangle \\
   &&\ \  +\bigl \langle  Z(t + (k + \alpha-1)\Delta t) \cdot Z(t' + (k + \beta-1)\Delta t) \cdot W(t'' + (k + \gamma)\Delta t)   \bigr \rangle \\
    &&\ \  - \bigl \langle  Z(t + (k + \alpha-1)\Delta t) \cdot Z(t' + (k + \beta-1)\Delta t) \cdot W(t'' + (k + \gamma-1)\Delta t)   \bigr \rangle \\
  &=&  B \cdot g(\alpha, \beta, \gamma, t, t', t''),\\
&&\mathrm{where}\\
&&g(\alpha, \beta, \gamma, t, t', t'')  \\
&=& \mathrm{min}\bigl(t+\alpha\Delta t,t'+\beta\Delta t, t''+ \gamma\Delta t\bigr) - \mathrm{min}\bigl(t+\alpha\Delta t,t'+\beta\Delta t, t''+(\gamma-1)\Delta t\bigr)\\
&-& \mathrm{min}\bigl(t+\alpha\Delta t,t'+(\beta-1)\Delta t, t''+\gamma\Delta t\bigr) + \mathrm{min}\bigl(t+\alpha\Delta t,t'+(\beta-1)\Delta t, t''+(\gamma -1)\Delta t\bigr)\\
&-& \mathrm{min}\bigl(t+(\alpha-1)\Delta t,t'+\beta\Delta t, t''+\gamma\Delta t\bigr) + \mathrm{min}\bigl(t+(\alpha-1)\Delta t,t'+\beta\Delta t, t''+(\gamma-1)\Delta t\bigr)\\
&+& \mathrm{min}\bigl(t+(\alpha-1)\Delta t,t'+(\beta-1)\Delta t, t''+\gamma\Delta t\bigr) - \mathrm{min}\bigl(t+(\alpha-1)\Delta t,t'+(\beta-1)\Delta t, t''+(\gamma-1)\Delta t\bigr).
\end{eqnarray*}
$g(\alpha, \beta, \gamma, t, t', t'') $ can be calculated to be
\begin{eqnarray*}
g(0, 0, 0, t, t', t'') &=& \Delta t - (t + t' +t'') + \mathrm{min}(t', t'' )  + \mathrm{min}(t, t'' )  + \mathrm{min}(t, t' ) \\
g(1, 0, 0, t, t', t'') &=& \mathrm{min}(t', t'') - \mathrm{min}(t, t', t'')\\
g(0, 1, 0, t, t', t'') &=& \mathrm{min}(t, t'') - \mathrm{min}(t, t', t'') \\
g(0, 0, 1, t, t', t'') &=& \mathrm{min}(t, t') - \mathrm{min}(t, t', t'') \\
g(1, 1, 0, t, t', t'') &=& t'' - \mathrm{min}(t, t'' )  - \mathrm{min}(t', t'')  + \mathrm{min}(t, t', t'' )  \\
g(1, 0, 1, t, t', t'') &=& t' - \mathrm{min}(t, t' )  - \mathrm{min}(t', t'')  + \mathrm{min}(t, t', t'' )  \\
g(0, 1, 1, t, t', t'') &=& t - \mathrm{min}(t, t' )  - \mathrm{min}(t, t'')  + \mathrm{min}(t, t', t'' ) .
\end{eqnarray*}
Thus,
\begin{equation*}
        \sum_{\substack{\alpha, \beta, \gamma=0\ \mathrm{or\ }1 \\ \mathrm{not\ }\alpha=\beta=\gamma= 1}} g(\alpha, \beta, \gamma, t, t', t'') = \Delta t
\end{equation*}

The error term 
\begin{equation*}
    \Omega(\alpha, \beta, \gamma) =  \Bigl\langle ( \varepsilon_{k+\alpha+1} - \varepsilon_{k+\alpha}) ( \varepsilon_{k+\beta+1} - \varepsilon_{k+\beta}) ( \delta_{k+\gamma+1} - \delta_{k+\gamma}) \Bigr\rangle
\end{equation*}
is calculated to be
\begin{eqnarray*}
\Omega(0, 0, 0) &=& 0 \\
\Omega(1, 0, 0) &=& \Omega(0, 1, 0) = \Omega(0, 0, 1) = -\langle \varepsilon_k^2 \delta_k \rangle \\
\Omega(0, 1, 1) &=& \Omega(1, 0, 1) = \Omega(1, 1, 0) = \langle \varepsilon_k^2 \delta_k \rangle.
\end{eqnarray*}
Thus,
\begin{equation*}
        \sum_{\substack{ \alpha, \beta, \gamma=0\ \mathrm{or\ }1\\ \mathrm{not\ }\alpha=\beta=\gamma= 1}} \Omega(\alpha, \beta, \gamma) = 0
\end{equation*}
Finally, we obtain
\begin{eqnarray}
\label{cve_B}
 &&\sum_{\substack{ \alpha, \beta, \gamma=0\ \mathrm{or\ }1\\ \mathrm{not\ }\alpha=\beta=\gamma= 1}} \langle \Delta Z_{k+\alpha} \cdot  \Delta Z_{k+\beta}  \cdot \Delta W_{k+\gamma} \rangle \nonumber \\
 && \ \ \ \ \ \ \ \ \ \ \ \ \ \ \ \ \ \   =  \int_0^{\Delta t} \int_0^{\Delta t} \int_0^{\Delta t} s(t)s(t')s(t'') B \Delta t dt dt' dt '' = B\Delta t.
\end{eqnarray}
Similarly, 
\begin{equation}
\label{cve_C}
    \sum_{\substack{ \alpha, \beta, \gamma=0\ \mathrm{or\ }1\\ \mathrm{not\ }\alpha=\beta=\gamma= 1}} \langle \Delta Z_{k+\alpha} \cdot  \Delta W_{k+\beta}  \cdot \Delta W_{k+\gamma} \rangle = C\Delta t
\end{equation}

\subsection{\label{sec:level2e}Covariance-based estimators for $E$}
$Z$ and $W$ are linked to $E$ by
\begin{equation*}
    \langle Z(t)^2 W(t)^2 \rangle - 2\langle Z(t)W(t)\rangle ^2 - \langle Z(t)^2 \rangle \langle W(t)^2\rangle = Et
\end{equation*}
With $f(x, y, z, w) = \langle xyzw \rangle - \langle xy\rangle \langle zw\rangle 
    - \langle xz\rangle \langle yw\rangle  - \langle xw\rangle \langle yz\rangle
$, this relation is written
\begin{equation*}
    f(Z(t), Z(t), W(t), W(t)) = Et
\end{equation*}
Thus, the relation that gives the covariance-based estimator of $E$ is predicted to be
\begin{equation}
\label{cve_E}
    \sum_{\substack{\alpha, \beta, \gamma, \lambda =0\ \mathrm{or\ }1 \\ \mathrm{not\ }\alpha=\beta=\gamma=\lambda= 1}} f( \Delta Z_{k+\alpha},  \Delta Z_{k+\beta}  , \Delta W_{k+\gamma}, \Delta W_{k+\lambda} ) = E\Delta t
\end{equation}

This can be proven by some algebra similar to the above proofs for the CVEs of $A$, $B$ and $C$  (see appendix A for details), confirming that the recipe for constructing CVEs (\ref{cve_recipe}) holds for the necessary cases to infer the anisotropic hopping rates based on the relation (\ref{pars_rates_relation}) (Fig. \ref{fig:rates_pars}).

\subsection{\label{sec:level2f}Inference of the hopping rates from the 2D trajectories simulated with motion blur and localization error}
\begin{figure*}[htb!]
\includegraphics[width=0.85\textwidth, angle=0]{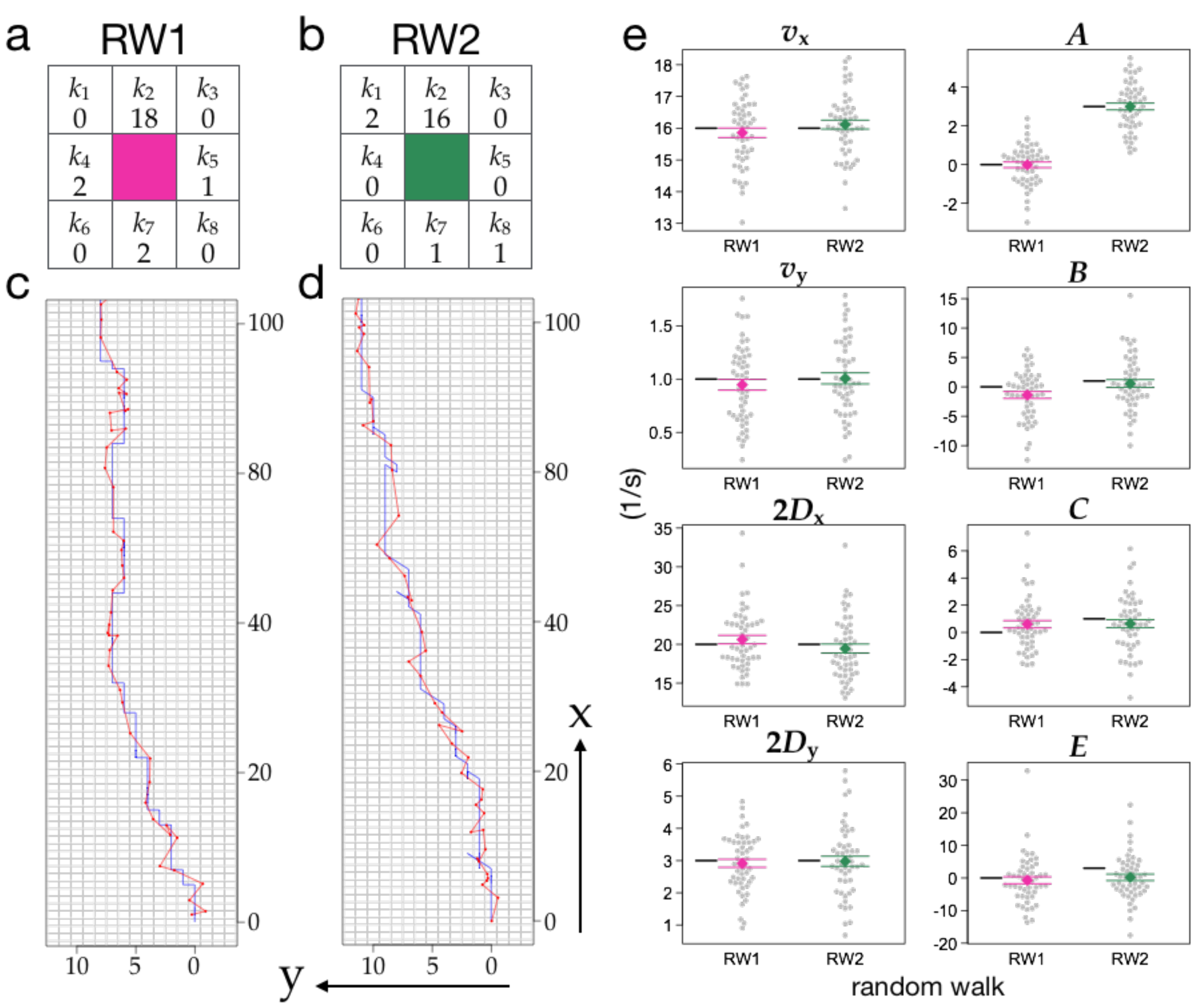}
\caption{\label{fig:trjs_pars} Monte Carlo simulation of anisotropic 2D random walks observed with motion blur and localization errors.\\ (a, b) Two random walks, RW1 and RW2, with the same x- and y- velocities and diffusion coefficients were generated based on the different sets of hopping rates. (c, d) Examples of simulated `true' positions on a 2D lattice (blue) and the `observed' trajectories (red) were shown. (e) Summary of the macroscopic parameters for the simulated RW1 and RW2 (magenta and green, respectively). For each random walk, 50 trajectories each consisting of 200 observations of the 0.1 s interval were simulated, and the CVEs of the parameters were calculated for each trajectory (gray dots). Means and standard errors are indicated in magenta and green. Black segments indicate the values expected by the theory (formula (\ref{pars_rates_relation}))  }
\end{figure*}

We now examine the versatility of the covariance-based estimators of $D_x, D_y, A, B, C,\mathrm{and\ } E $ (formulas \ref{cve_2Dx}, \ref{cve_2Dy},  \ref{cve_A}, \ref{cve_B}, \ref{cve_C}, and \ref{cve_E}), the coefficients of the time development of the co-moments of the displacements, and whether we can infer the anisotropic hopping rates from the trajectory data of 2D random walks, using the relations that link the macroscopic coefficients to the microscopic hopping rates (formula \ref{pars_rates_relation}). Let us first compare two examples of random walks, RW1 and RW2, respectively, that are generated by Monte Carlo simulations to have the same x- and y-velocities and diffusion coefficients ($v_x, v_y, D_x, D_y$), but have different hopping rates (Fig. \ref{fig:trjs_pars} a and b). The values of the velocities, $v_x = 16\  \mathrm{s}^{-1}$ and $v_y = 1\ \mathrm{s}^{-1}$ (hops per second) correspond, for example, to the helical motion of a kinesin-like motor protein with a helical pitch of $\sim$1.7 $\mu\mathrm{m}$ around a microtubule that has a 2D lattice of the discrete motor-binding sites on its surface consisting of parallel 13 protofilaments, i.e., linear arrays of tubulin subunits aligned at 8 nm periodicity \cite{Chretien:1991ct, Wade:1993ib}. To mimic a typical condition of image acquisition using an EM-CCD camera used for the single molecule/particle observation, a series of ‘true’ positions are generated by the Gillespie algorithm \cite{Gillespie:1977ww} (blue lines in Fig. \ref{fig:trjs_pars} c and d) and the ‘observed’ positions at regular time points (every 0.1 s for 200 time points, corresponding to observation for 20 sec) were calculated by averaging the positions during the open shutter (90\% of the cycle) and by adding Gaussian noises of the standard deviation of half the size of the 
lattice spacing (red lines in Fig. \ref{fig:trjs_pars} c and d).

Fifty trajectories of RW1 and RW2 were generated, respectively, and the macroscopic parameters $v_x, v_y, D_x, D_y, A, B, C,\mathrm{and\ } E $ for each trajectory were calculated based on the formulas \ref{cve_vx}, \ref{cve_vy}, \ref{cve_2Dx}, \ref{cve_2Dy}, \ref{cve_A}, \ref{cve_B}, \ref{cve_C}, and \ref{cve_E}. As shown in Fig. \ref{fig:trjs_pars} e, the covariance-based estimators calculated from the simulated trajectories (gray dots) distributed around the theoretical values predicted by the formula (\ref{pars_rates_relation}) (solid lines). Although, due to the intrinsic stochasticity of the random walk, the calculated values from individual trajectories spread around the theoretical `true' values, the means of 50 trajectories were found close to the corresponding theoretical values within twice the standard error. Importantly, as expected, while there was no significant difference between RW1 and RW2 in $v_x, v_y, D_x, \mathrm{and\ } D_y$, a clear difference between them was found in the distribution of $A$. This suggests that our recipe gives reasonable estimates of the coefficients of the temporal development of the co-moments of the x- and y-displacements in a realistic scenario. This allows us to detect a difference between the two random walks that have distinct preferences in the hopping direction but look exactly the same if we only consider the velocities and the diffusion coefficients along the two axes, separately.

\begin{figure*}[htb!]
\includegraphics[width=0.86\textwidth, angle=0]{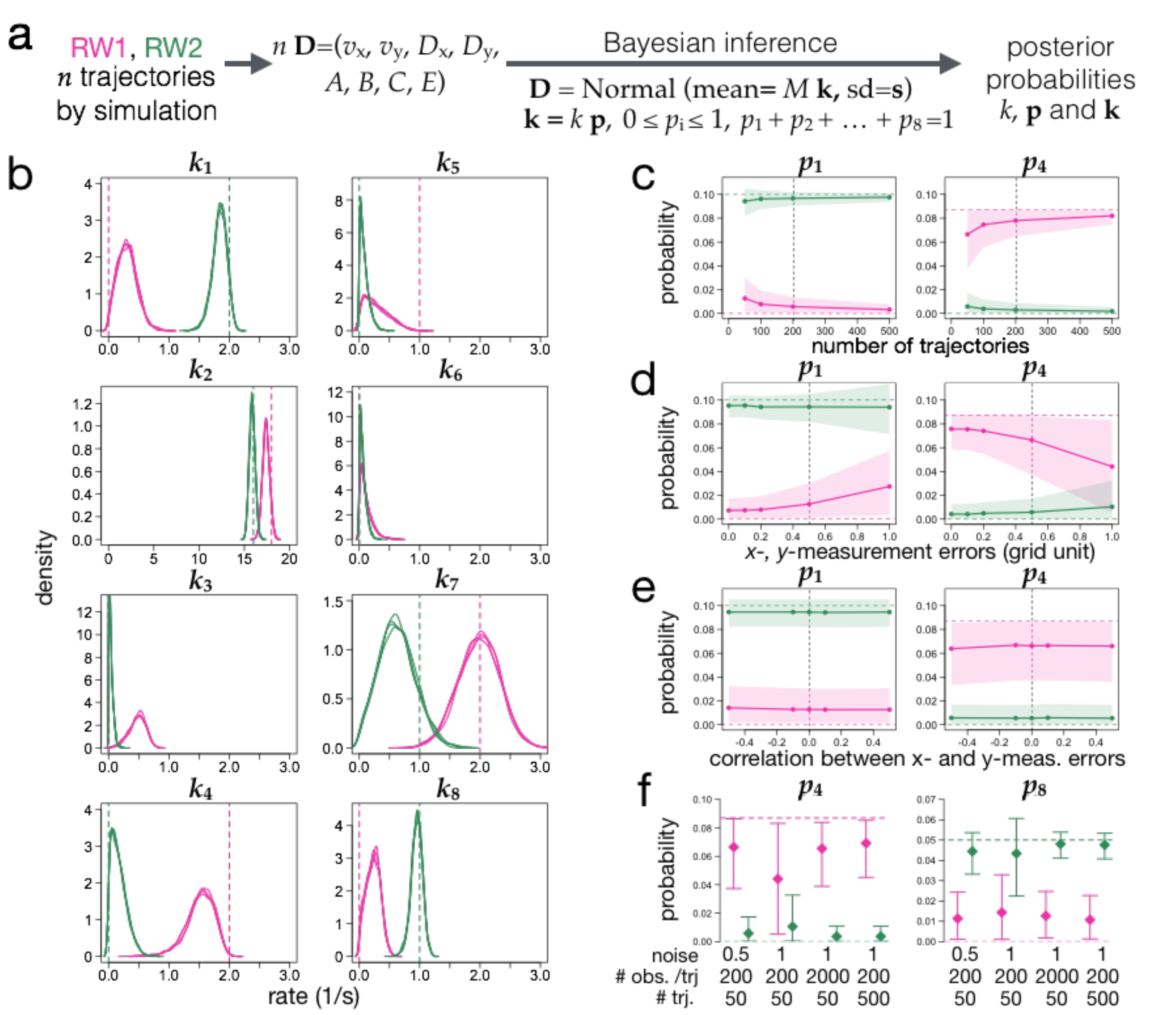}
\caption{\label{fig:rates_probs} Inference of the hopping rates and hopping preferences.\newline 
(a) Procedure of Bayesian inference of the hopping rates and hopping preferences from the simulated trajectories. Covariance-based estimators of the macroscopic coefficients ($\textbf{D}$) calculated for individual trajectories were used as data to compute the posterior probabilities of the hopping rates ($\textbf{k}$) and hopping preferences  ($\textbf{p}$), which are linked to each other via the overall hopping rate ($k$), based on a model that the observed macroscopic coefficients are probabilistic variables that distribute around the theoretical values, $M \textbf{k}$, where $M$ is the matrix that appears in formula \ref{pars_rates_relation}. (b) Posterior probability distributions of the hopping rates (RW1:magenta, RW2:green). Curves represent the results of the four independent chains of Bayesian inference. Dashed lines indicate the theoretical values. (c to e) Influences of the data size (c), localization errors (d) and the correlation between the x- and y-localization errors (e). Dots linked with solid lines and shaded regions represent the averages of the means and 95\% credible intervals of 100 independent trials, respectively. Horizontal dashed lines are true values. Vertical dashed lines indicate the parameter values used in the simulation in (b). (f) Loss of the estimation accuracy by increased localization errors and its recovery by expansion of the data size (the number of the observations per trajectory or the number of the trajectories). Means and 95\% credible intervals.
}
\end{figure*}

We move on to examine the power of our approach to reveal the hopping behaviors of anisotropic random walks. With known macroscopic parameters $(v_x, v_y, D_x, D_y, A, B, C, E)$, we can calculate $(k_1, k_2, k_3, k_4, k_5, k_6, k_7, k_8)$, the hopping rates to the 8 surrounding sites, by solving by the formula (\ref{pars_rates_relation}). However, due to the intrinsically stochastic behavior of the random walks, a simple solution might result in negative values of the hopping rates. To avoid this, we performed a Bayesian inference with a model that the observed macroscopic parameters are probabilistic variables that distribute around the true values that are defined by the microscopic hopping rates, which don't take negative values (Fig. \ref{fig:rates_probs} a). This was implemented with Stan \cite{Rstan:2020} on R \cite{R_core_team:2019} with the overall hopping rate, $k$ as a positive real number and the hopping preferences $\bf{p}$ $=(p_1, p_3, p_4, p_5, p_6, p_7, p_8)$ as simplex variables ($0\leq p_i \leq 1, \sum p_i =1$), where the individual hopping rates $k_i = k \cdot p_i$ with $k=\sum_i k_i$. As priors, uniform probabilities were used.

Fig. \ref{fig:rates_probs} b shows the distributions of the posterior probabilities of the hopping rates, obtained with the fifty sets of macroscopic parameters $(v_x, v_y, D_x, D_y, A, B, C, E)$ for each Brownian motion from Fig. \ref{fig:trjs_pars} e as the data for the Bayesian inference. As expected, RW1 (magenta) and RW2 (green) showed clearly distinct patterns in the hopping preferences, which are consistent with the theoretical values used for generation of the random walks by simulation. For example, the (posterior) probability of the hopping rate to the forward-left, $k_1$, of RW1 showed a distribution near the theoretical value $0\ \mathrm{s^{-1}}$ while that of RW2 had a peak near $2\ \mathrm{s^{-1}}$. This indicates that our approach can properly infer the microscopic parameters to the precision levels that are sufficient to distinguish the two example cases of random walk, which would look the same if we only analyze the movements in x- and y-directions separately.

If we look closer, we realize that the peak positions of the posterior probabilities of the hopping rates don't exactly match with the theoretical values. This is likely due to the probabilistic uncertainties in the observed random walks, which are influenced by the length and the number of the trajectories and by the precision of the measurements (the motion blurs and localization errors), in combination with the natural condition that the hopping rates can't be less than zero. To assess the effects of the measurement conditions, the cycle of the generation of the observed trajectories by simulation, estimation of the coefficients for each trajectory, and inference of the rates and preferences of hopping was repeated a hundred times per condition. The average of the means and 95\% credible intervals of the posterior distribution of the 100 independent trials were averaged and shown in Fig. \ref{fig:rates_probs} c to e and Supplemental Figures.

As expected, increasing the size of data by increasing the number of trajectories sharpened the distribution of the posterior probabilities of the rates and preferences of hopping (Fig. \ref{fig:rates_probs} c and Supplementary Figure 1). Importantly, their means, which showed deviations from the theoretical values when the number of input trajectories was small as mentioned above, asymptotically approached their respective theoretical values as the available trajectories increase. Similar sharpening of the distribution and approaching to the theoretical value were observed also when the length of each trajectory was increased instead (Supplementary Figure 2).

The localization errors don't explicitly appear in the formulas of the CVEs (\ref{cve_vx}, \ref{cve_vy}, \ref{cve_2Dx}, \ref{cve_2Dx}, \ref{cve_A}, \ref{cve_B}, \ref{cve_C}, and \ref{cve_E}) because we took averages of infinitely many combinations of the displacements. With the finite displacement data, however, the noise terms would not be completely canceled out and affect the estimations. Indeed, increasing the amplitude of the localization errors significantly broadens the posterior probability distributions of the hopping preferences (and the hopping rates) (Fig. \ref{fig:rates_probs} d and Supplementary Figure 3). For example, with the data of the 50 trajectories with 200 observations each, although the $p_4$ values of RW1 and RW2 were distinguishable in the presence of the localization errors up to half the size of the lattice unit, they became indistinguishable when the amplitude of the errors was increased to be the same as the lattice unit (Fig. \ref{fig:rates_probs} d). This is in contrast with little impact of the extent of correlation between the x- and y-measurement errors for each observation  (Fig. \ref{fig:rates_probs} e, Supplementary Figure 4). Interestingly, the perturbation by the localization errors could be overcome by increasing the size of data, either by increasing the number of observations per trajectory or by observing more trajectories (Fig. \ref{fig:rates_probs} f). The difference between RW1 and RW2 that was once obscured by the increased localization errors (1 grid size) became distinguishable again with a 10-fold increase in the amount of data either by increasing the number of observations made per trajectory or by increasing the number of trajectories observed.

\begin{figure*}[htb!]
\includegraphics[width=0.9 \textwidth, angle=0]{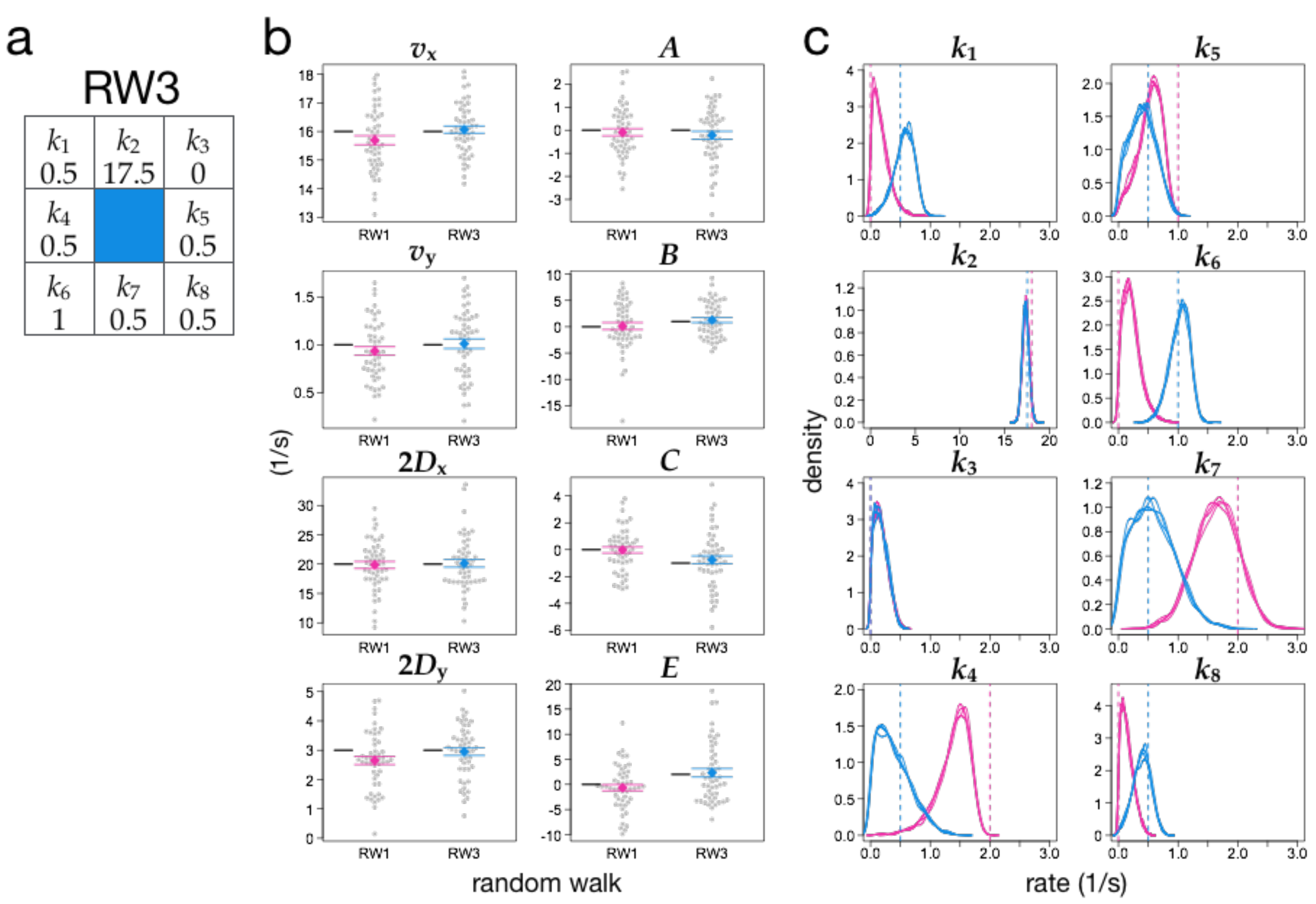}
\caption{\label{fig:rw3} Further test of our approach with RW3, a random walk with theoretical values of $v_x, v_y, D_x, D_y, \mathrm{and\ } A$ identical to those of RW1. \newline (a) Hopping rates of random walk, RW3, designed to result in the same theoretical values of $v_x, v_y, D_x, D_y, \mathrm{and\ } A$ as those of RW1. (b) Macroscopic coefficients of RW1 and RW3 calculated from the 50 trajectories, respectively, each with 200 observations generated by Monte Carlo simulation. The means and standard errors (magenta and blue, respectively) are shown with the values for individual trajectories (gray dots). Black segments indicate the theoretical values by the formula (\ref{pars_rates_relation}). (c) Distributions of the posterior probabilities of the hopping rates inferred with the coefficients in (b). Curves represent the results of four independent chains performed for each random walk. Dashed vertical lines indicate the theoretical values.}
\end{figure*}

Although the hopping rates of RW1 and RW2 were set to have identical theoretical values of $v_x, v_y, D_x, \mathrm{and\ } D_y$, their theoretical values of $A$ were different (0 vs 3 $\mathrm{s^{-1}}$). Thus, we can't exclude the possibility that the above success in discriminating between the hopping patterns of RW1 and RW2 might solely rely on the distinct values of $A$, irrespective of the  coefficients for the other co-moments. To test whether our approach can distinguish two random walks that have identical $v_x, v_y, D_x, D_y, \mathrm{and\ }A$, we finally consider another random walk, RW3 (Fig. \ref{fig:rw3} a), whose theoretical $A$, in addition to the theoretical values of $v_x, v_y, D_x, \mathrm{and\ } D_y$, is identical to that of RW1. As expected, the coefficients $v_x, v_y, D_x, D_y, \mathrm{and\ } A$ calculated from the 50 RW3 trajectories each with 200 observations showed distributions indistinguishable from the corresponding ones from the RW1 trajectories  (Fig. \ref{fig:rw3} b). In contrast, the distributions of the calculated $C$ and $E$ values of RW3 were distinct from those of RW1. As expected, the posterior probabilities of the hopping rates of RW3 closely reproduced the values set for the simulation of the trajectories, exhibiting a difference from the corresponding one of RW1 ($k_1, k_4, k_6, \mathrm{and\ }k_7$) (Fig. \ref{fig:rw3} c). This suggests that our approach provides reasonable inference of the hopping behaviors under a realistic setting for the single particle observation even in a case where $B, C$ or $E$, the coefficients for the higher-order co-moments of the drift-adjusted x- and y-displacements, are the sole observable clues.

\section{Discussion}
Here we studied the anisotropic random walk on a 2D lattice and derived the relationship between the macroscopic coefficients of the time-development of the drift-adjusted displacements and the microscopic hopping rates. We then extended the covariance-based estimators of the 1D diffusion coefficient to the higher-order co-moments of the drift-adjusted displacements and used them to infer the hopping rates from the trajectory data affected by motion blur and localization errors. The versatility of this novel approach was evaluated with the trajectory data generated by simulation. We could demonstrate that our approach can distinguish 2D random walks that have exactly the same x- and y-velocities and diffusion coefficients but have distinct hopping patterns. 

An important advantage of the covariance-based estimators \cite{Vestergaard:2016ko,Vestergaard:2015jx,Vestergaard:2014bk} is that the equations to calculate them don't explicitly contain terms for localization errors nor motion blur after taking the average of infinite terms of observed x- and y-displacements combined. This assumption, of course, is not true with the real-world data of finite size. With a finite size of data, the failure in cancellation of the interfering factors as well as the intrinsic uncertainly of the random process results in an error in the estimation of the coefficients that characterize the random walk. Although it had been demonstrated that the CVE-based approach is superior to the traditional ones in the case of 1D unbiased diffusion \cite{Vestergaard:2016ko,Vestergaard:2015jx,Vestergaard:2014bk}, it was unclear how robust a 2D version of the CVE-based approach would be since it is involved with the higher-order terms of the displacements. However, through the analysis of the simulated model cases of 2D random walk, it was confirmed that the CVE-based approach combined with Bayesian inference can reasonably estimate the anisotropic hopping patterns, with a realistic size of data when the precision of the measurement is smaller than the size of the grid. Even with a lower precision of the measurements, increasing the size of data could restore the accuracy of the estimation. For real-world data, careful optimization will be necessary, considering rather complicated and intertwined influences of the frequency of image acquisition, exposure time, the intensity of illumination \cite{Shen:2017dq,Hoze:2017wa,Vestergaard:2016ko,Manzo:2015dc,Chenouard:2014kg}. 

It has recently been reported that some kinesin-like motor proteins show a helical motion around a microtubule \cite{Bormuth:2012kc, Mitra:2019eq, Bugiel:2018df, Mitra:2018bn}. The tubular structure of a microtubule consists of 13 or 14 protofilaments, i.e., the linear arrays of $\alpha$- and $\beta$-tubulin heterodimers with the 8 nm periodicity, presenting the motor-binding sites as the 2D lattice on the surface \cite{Chretien:1991ct, Wade:1993ib}. Interestingly, the reported pitches of the helical are intermediate between the shortest helical pitch of the lattice due to the staggered alignment of the protofilaments and the longest helical pitch observed in the 14-protofilament tubule. The observed helical motion corresponds to the protofilament switch that occurs once per $\sim$10 forward steps on average, implying the stochastic stepping patterns to the neighboring sites. Our approach might be applicable to estimate such patterns based on the observed helical trajectories.

On a 3D lattice, there are $26 (=9+8+9)$ choices for stochastic hopping to a nearby site. The relation between the macroscopic coefficients and the hopping rates for a 3D lattice analogous to the 2D version (\ref{pars_rates_relation}) will contain 26 hopping rates and 26 coefficients for the time-development of the combinations of (higher-order co-)moments of the drift-adjusted displacements up to the 6th order. Whether the recipe to derive the covariance-based estimators of the coefficients for the higher-order co-moments developed here for a 2D lattice (\ref{cve_recipe}) is applicable to a 3D lattice will be a future question.

\begin{acknowledgments}
The author would like to thank Huong T. Vu (University of Warwick, UK), Matthew Turner (University of Warwick, UK), Junichiro Yajima (University of Tokyo, Japan) and Izumu Mishima (Durham University, UK) for insightful discussions and critical reading of the manuscript.
\end{acknowledgments}

\bibliography{2dcve}

\newpage
\appendix

\section{Proof of the equation for the covariance-based estimator of $E$}
Here we will prove Equation \ref{cve_E}.
Let us define notations, $\Delta \widehat{Z}(t, k, \alpha) = Z(t+(k+\alpha)\Delta t) - Z(t + (k+\alpha -1)\Delta t)$ and $ \Delta \varepsilon_{k, \alpha} =\varepsilon_{k+\alpha+1} - \varepsilon_{k+\alpha}$. With these, 
\begin{eqnarray}
&& f( \Delta Z_{k+\alpha},  \Delta Z_{k+\beta}  , \Delta W_{k+\gamma}, \Delta W_{k+\lambda} ) \nonumber \\
&=& \Bigl \langle  
    \Bigl[\int_0^{\Delta t} s(t) \Delta \widehat{Z}(t, k, \alpha)dt + \Delta \varepsilon_{k, \alpha}\Bigr] 
    \times \Bigl[\int_0^{\Delta t} s(t') \Delta \widehat{Z}(t', k, \beta)dt' + \Delta \varepsilon_{k, \beta}\Bigr] \nonumber \\
&&    \times \Bigl[ \int_0^{\Delta t} s(t'') \Delta \widehat{W}(t'', k, \gamma)dt'' + \Delta \delta_{k, \gamma}\Bigr]  
    \times  \Bigl[ \int_0^{\Delta t} s(t''') \Delta \widehat{W}(t''', k, \lambda)dt''' + \Delta \delta_{k, \lambda}\Bigr] 
\Bigr \rangle \nonumber\\
&& \ \ \ \ \  - \Bigl \langle  \Bigl[\int_0^{\Delta t} s(t) \Delta \widehat{Z}(t, k, \alpha)dt + \Delta \varepsilon_{k, \alpha}\Bigr]  \times \Bigl[\int_0^{\Delta t} s(t') \Delta \widehat{Z}(t', k, \beta)dt' + \Delta \varepsilon_{k, \beta}\Bigr] \Bigr \rangle \nonumber\\
&& \ \ \ \ \ \  \ \ \ \ \  \times \Bigl \langle \Bigl[ \int_0^{\Delta t} s(t'') \Delta \widehat{W}(t'', k, \gamma)dt'' + \Delta \delta_{k, \gamma}\Bigr] \times \Bigl[ \int_0^{\Delta t} s(t''') \Delta \widehat{W}(t''', k, \lambda)dt''' + \Delta \delta_{k, \lambda}\Bigr]  \Bigr \rangle \nonumber\\
&& \ \ \ \ \ - \Bigl \langle \Bigl[\int_0^{\Delta t} s(t) \Delta \widehat{Z}(t, k, \alpha)dt + \Delta \varepsilon_{k, \alpha}\Bigr] \times \Bigl[ \int_0^{\Delta t} s(t'') \Delta \widehat{W}(t'', k, \gamma)dt'' + \Delta \delta_{k, \gamma}\Bigr]  \Bigr \rangle \nonumber\\
&& \ \ \ \ \ \  \ \ \ \ \  \times \Bigl \langle  \Bigl[\int_0^{\Delta t} s(t') \Delta \widehat{Z}(t', k, \beta)dt' + \Delta \varepsilon_{k, \beta}\Bigr] \times \Bigl[ \int_0^{\Delta t} s(t''') \Delta \widehat{W}(t''', k, \lambda)dt''' + \Delta \delta_{k, \lambda}\Bigr] \Bigr \rangle \nonumber\\
&& \ \ \ \ \  - \Bigl \langle \Bigl[\int_0^{\Delta t} s(t) \Delta \widehat{Z}(t, k, \alpha)dt + \Delta \varepsilon_{k, \alpha}\Bigr] \times \Bigl[ \int_0^{\Delta t} s(t''') \Delta \widehat{W}(t''', k, \lambda)dt''' + \Delta \delta_{k, \lambda}\Bigr] \Bigr \rangle \nonumber\\
&& \ \ \ \ \ \  \ \ \ \ \  \times \Bigl \langle  \Bigl[\int_0^{\Delta t} s(t') \Delta \widehat{Z}(t', k, \beta)dt' + \Delta \varepsilon_{k, \beta}\Bigr] \times \Bigl[ \int_0^{\Delta t} s(t'') \Delta \widehat{W}(t'', k, \gamma)dt'' + \Delta \delta_{k, \gamma}\Bigr]  \Bigr \rangle \nonumber\\
&&= \int_0^{\Delta t} \int_0^{\Delta t} \int_0^{\Delta t} \int_0^{\Delta t} s(t)s(t')s(t'') s(t''') \times \nonumber\\
&& \ \ \ \ \ \ \ \ \ \ \ \ f(\Delta \widehat{Z}(t, k, \alpha), \Delta \widehat{Z}(t', k, \beta), \Delta \widehat{W}(t'', k, \gamma) , \Delta \widehat{W}(t''', k, \lambda)  ) dt dt' dt'' dt ''' \nonumber\\ \label{integral_errors_decomposed}
&& \ \ \ \ \ \  \ \ \ \ \ \ \ \ \ \ \ \ \ \ \ \ \ \ +  f(\Delta \varepsilon_{k, \alpha}, \Delta \varepsilon_{k, \beta},  \Delta \delta_{k, \gamma}, \Delta\delta_{k, \lambda})
\end{eqnarray}
In the above transformation, we used that the true displacements and the measurement errors are independent of each other and that the ensemble means of the true displacements and those of the measurement errors are both zero. 

Now we evaluate $f(\Delta \widehat{Z}(t, k, \alpha), \Delta \widehat{Z}(t', k, \beta), \Delta \widehat{W}(t'', k, \gamma) , \Delta \widehat{W}(t''', k, \lambda)  )$ in the first term. For unbiased random walks $A(s), B(t), C(u), D(v)$ ($\langle A(s) \rangle=\langle B(t) \rangle=\langle C(u) \rangle = \langle D(v) \rangle = 0$, and their displacements are independent for non-overlapping time sections), we can show, by a similar calculation to (\ref{wiener_process}), that
\begin{equation}
    f(A(s), B(t), C(u), D(v)) = f(A(\tau), B(\tau), C(\tau), D(\tau)),
\end{equation}
where $\tau = \mathrm{min}(s, t, u, v)$. Thus, with $\tau = \mathrm{min}(t +(k+\alpha -i)\Delta t, t' +(k+\beta -i')\Delta t, t'' +(k+\gamma -i'')\Delta t,  t''' +(k+\lambda -i''')\Delta t)$,
\begin{eqnarray*}
    && f(\Delta \widehat{Z}(t, k, \alpha), \Delta \widehat{Z}(t', k, \beta), \Delta \widehat{W}(t'', k, \gamma) , \Delta \widehat{W}(t''', k, \lambda)  )\\
    &=& \sum_{i=0, 1} \sum_{i'=0, 1} \sum_{i''=0, 1} \sum_{i'''=0, 1} (-1)^{i + i' + i'' + i'''} \times  \\
    && f(Z(t +(k+\alpha -i)\Delta t), Z(t' +(k+\beta -i')\Delta t), W(t'' +(k+\gamma -i'')\Delta t), W(t''' +(k+\lambda -i''')\Delta t)) \\
    &=& \sum_{i=0, 1} \sum_{i'=0, 1} \sum_{i''=0, 1} \sum_{i'''=0, 1} (-1)^{i + i' + i'' + i'''}  f(Z(\tau), Z(\tau), W(\tau), W(\tau)) \\
    &=& \sum_{i=0, 1} \sum_{i'=0, 1} \sum_{i''=0, 1} \sum_{i'''=0, 1} (-1)^{i + i' + i'' + i'''}  E\cdot \tau \\
    &=& E \cdot \sum_{i=0, 1} \sum_{i'=0, 1} \sum_{i''=0, 1} \sum_{i'''=0, 1} (-1)^{i + i' + i'' + i'''} \times \\
    &&\ \ \ \ \ \mathrm{min}\bigl[ +(k+\alpha -i)\Delta t, t' +(k+\beta -i')\Delta t, t'' +(k+\gamma -i'')\Delta t,  t''' +(k+\lambda -i''')\Delta t\bigr]
\end{eqnarray*}
Here, let us consider
\begin{eqnarray*}
&& g(\alpha, \beta, \gamma, \lambda, t, t', t'', t''') = \sum_{i=0, 1} \sum_{i'=0, 1} \sum_{i''=0, 1} \sum_{i'''=0, 1} (-1)^{i + i' + i'' + i'''} \times \\
&&\ \ \ \ \ \mathrm{min}\bigl[ +(k+\alpha -i)\Delta t, t' +(k+\beta -i')\Delta t, t'' +(k+\gamma -i'')\Delta t,  t''' +(k+\lambda -i''')\Delta t\bigr].
\end{eqnarray*}

The values of 15 combinations of $\alpha, \beta, \gamma, \mathrm{and\ } \lambda = 0 \mathrm{\ or\ } 1$ (except for $\alpha=\beta=\gamma=\lambda = 1$) are evaluated as 
\begin{eqnarray*}
g(0, 0, 0, 0, t, t', t'', t''') &=& \Delta t - (t + t' + t'' + t''') \\
+ \mathrm{min}(t, t') & +& \mathrm{min}(t, t'')  + \mathrm{min}(t, t''') + \mathrm{min}(t', t'') + \mathrm{min}(t', t''')  + \mathrm{min}(t'', t''') \\
- \bigl[ \mathrm{min}(t', t'', t''') &+& \mathrm{min}(t, t'', t''') + \mathrm{min}(t, t', t''') + \mathrm{min}(t, t', t'')\bigr ] +2 \cdot \mathrm{min}(t, t', t'', t''') \\
g(1, 0, 0, 0, t, t', t'', t''') &=& \mathrm{min}(t', t'', t''') - \mathrm{min}(t, t', t'', t''')\\
g(1, 1, 0, 0, t, t', t'', t''') &=& \mathrm{min}(t'', t''') - \mathrm{min}(t', t'', t''') - \mathrm{min}(t, t'', t''') + \mathrm{min}(t, t', t'', t''')\\
g(1, 1, 1, 0, t, t', t'', t''') &=& t''' - \bigl[ \mathrm{min}(t, t''') + \mathrm{min}(t', t''') + \mathrm{min}(t'', t''', t, t', t'', t''') \bigr] \\
 &+& \mathrm{min}(t, t', t''') + \mathrm{min}(t, t'', t''') + \mathrm{min}(t', t'', t''')  - \mathrm{min}(t, t', t'', t'''),
\end{eqnarray*}
etc.

All the terms except for $\Delta t$ cancel out (Table.~\ref{table1:g_values}).

\begin{table}[ht]
\caption{\label{table1:g_values} Calculation of $g(\alpha, \beta, \gamma, \lambda, t, t', t'', t''')$. Counts of the consisting terms for various combinations of $\alpha, \beta, \gamma,\mathrm{and\ } \lambda.$}
\begin{tabular}{|l|l|l|l|l|}
\hline
                     & $t, t'$, etc. & min($t, t'$), etc. & min($t, t', t''$), etc. & min($t, t', t'', t'''$) \\ \hline
$g(0, 0, 0, 0, t, t', t'', t''')$        & $-1$ each 4 & $+1$ each 6    & $-1$ each 4       & $+2$            \\ \hline
$g(1, 0, 0, 0, t, t', t'', t''')$ etc. 4 &             &                & $+1$ each 4       & $-1$ each 4     \\ \hline
$g(1, 1, 0, 0, t, t', t'', t''')$ etc. 6 &             & $+1$ each 6    & $-2$ each 6       & $+1$ each 6     \\ \hline
$g(1, 1, 1, 0, t, t', t'', t''')$ etc. 4 & $+1$ each 4 & $-3$ each 4    & $+3$ each 4       & $-1$ each 4     \\ \hline
sum                  & $-4+4=0$          & $6+6-12=0$  & $-4+4-12+12=0$ & $+2-4 +6-4 =0$          \\ \hline
\end{tabular}
\end{table}

Thus,
\begin{equation*}
    \sum_{\substack{\alpha, \beta, \gamma, \lambda =0\ \mathrm{or\ }1 \\ \mathrm{not\ }\alpha=\beta=\gamma=\lambda= 1}} g(\alpha, \beta, \gamma, \lambda, t, t', t'', t''')=\Delta t
\end{equation*}
i.e.,
\begin{equation}
\label{true_before_integration}
\sum_{\substack{\alpha, \beta, \gamma, \lambda =0\ \mathrm{or\ }1 \\ \mathrm{not\ }\alpha=\beta=\gamma=\lambda= 1}} f(\Delta \widehat{Z}(t, k, \alpha), \Delta \widehat{Z}(t', k, \beta), \Delta \widehat{W}(t'', k, \gamma) , \Delta \widehat{W}(t''', k, \lambda)  ) = E\Delta t
\end{equation}

The second term of $f( \Delta Z_{k+\alpha},  \Delta Z_{k+\beta}  , \Delta W_{k+\gamma}, \Delta W_{k+\lambda} )$ (the formula \ref{integral_errors_decomposed}) can be decomposed into a sum of the terms with various combinations of $i, i', i'', i'''$.
\begin{eqnarray*}
&&f(\Delta \varepsilon_{k, \alpha}, \Delta \varepsilon_{k, \beta},  \Delta \delta_{k, \gamma}, \Delta\delta_{k, \lambda}) \\
&=&\ \  \sum_{i=0, 1} \sum_{i'=0, 1} \sum_{i''=0, 1} \sum_{i'''=0, 1} (-1)^{i + i' + i'' + i'''}  f(\varepsilon_{k+\alpha+1-i}, \varepsilon_{k+\beta+1-i'}, \delta_{k+\gamma+1-i''}, \delta_{k+\lambda+1-i'''}) 
\end{eqnarray*}
We need to evaluate 15 combinations of $\alpha, \beta, \gamma, \mathrm{and\ } \lambda = 0 \mathrm{\ or\ } 1$ (except for $\alpha=\beta=\gamma=\lambda = 1$) each for 16 combinations of $i, i', i'',\mathrm{and\ } i'''$. Most of the $15\times 16$ terms are zero. As in  Table.~\ref{table1:f_errors}), the other non-zero terms are cancelled out each other. 

\begin{table}[ht]
\caption{\label{table1:f_errors} Calculation of $f(\Delta \varepsilon_{k, \alpha}, \Delta \varepsilon_{k, \beta},  \Delta \delta_{k, \gamma}, \Delta\delta_{k, \lambda})$. All the possible combinations of $\alpha$, $\beta$, $\gamma$, and $\lambda$ (excluding $\alpha=\beta=\gamma=\lambda=1$) that make $\langle \varepsilon_{k+\alpha -i} \varepsilon_{k+\beta-i'} \delta_{k+\gamma -i''} \delta_{k+\lambda-i'''} \rangle$ or $\langle \varepsilon_{k+\alpha -i} \varepsilon_{k+\beta-i'}\rangle \langle \delta_{k+\gamma -i''} \delta_{k+\lambda-i'''} \rangle$ non-zero for different combinations of $i, i', i'',\mathrm{and\ } i'''$.}
\includegraphics[width=1.0\textwidth, angle=0]{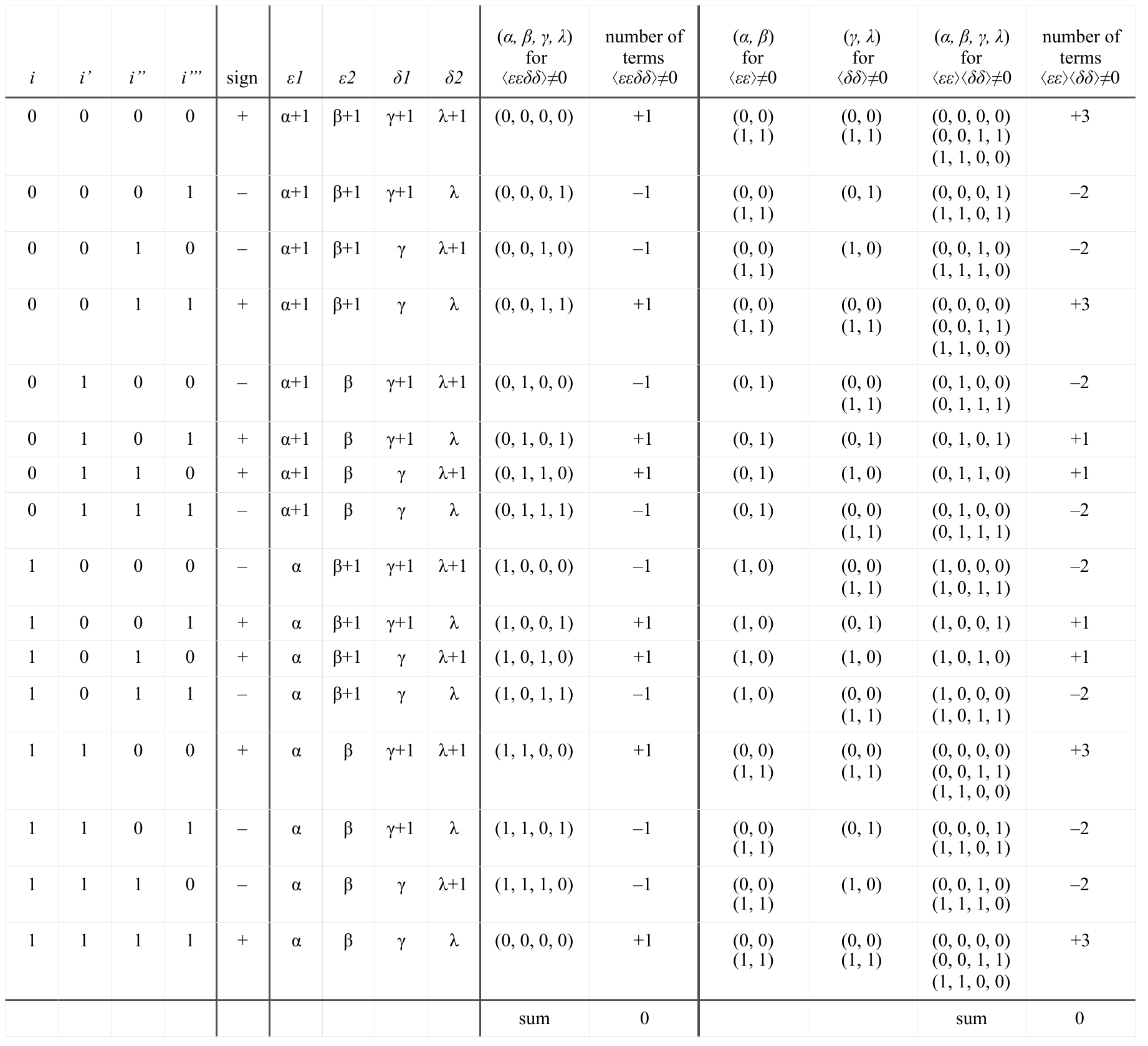}
\end{table}

Thus, 
\begin{equation}
\label{noerror}
    \sum_{\substack{\alpha, \beta, \gamma, \lambda =0\ \mathrm{or\ }1 \\ \mathrm{not\ }\alpha=\beta=\gamma=\lambda= 1}} f(\Delta \varepsilon_{k, \alpha}, \Delta \varepsilon_{k, \beta},  \Delta \delta_{k, \gamma}, \Delta\delta_{k, \lambda})=0
\end{equation}
Finally, by plugging (\ref{true_before_integration}) and (\ref{noerror}) into (\ref{integral_errors_decomposed}), 
\begin{eqnarray*}
    &&\sum_{\substack{\alpha, \beta, \gamma, \lambda =0\ \mathrm{or\ }1 \\ \mathrm{not\ }\alpha=\beta=\gamma=\lambda= 1}} f( \Delta Z_{k+\alpha},  \Delta Z_{k+\beta}  , \Delta W_{k+\gamma}, \Delta W_{k+\lambda} ) \\
    &=& \int_0^{\Delta t} \int_0^{\Delta t} \int_0^{\Delta t} \int_0^{\Delta t} s(t)s(t')s(t'') s(t''') \times \\
&& \ \ \ \ \ \ \sum_{\substack{\alpha, \beta, \gamma, \lambda =0\ \mathrm{or\ }1 \\ \mathrm{not\ }\alpha=\beta=\gamma=\lambda= 1}} f(\Delta \widehat{Z}(t, k, \alpha), \Delta \widehat{Z}(t', k, \beta), \Delta \widehat{W}(t'', k, \gamma) , \Delta \widehat{W}(t''', k, \lambda)  ) dt dt' dt'' dt ''' \\
&& \ \ \ \ \ \  \ \ \ \ \ \  + \sum_{\substack{\alpha, \beta, \gamma, \lambda =0\ \mathrm{or\ }1 \\ \mathrm{not\ }\alpha=\beta=\gamma=\lambda= 1}}  f(\Delta \varepsilon_{k, \alpha}, \Delta \varepsilon_{k, \beta},  \Delta \delta_{k, \gamma}, \Delta\delta_{k, \lambda}) \\
 &=& \int_0^{\Delta t} \int_0^{\Delta t} \int_0^{\Delta t} \int_0^{\Delta t} s(t)s(t')s(t'') s(t''') E\cdot \Delta t dt dt' dt'' dt''' = E\Delta t
\end{eqnarray*}
we obtain the formula (\ref{cve_E}).

\end{document}